\documentclass[aps,pre,reprint,twocolumn,showpacs,amsmath,amssymb]{revtex4-1}
\usepackage[T1]{fontenc}
\usepackage{amsmath}
\usepackage{graphicx}
\usepackage{bm}
\usepackage{color}
\usepackage{lineno}
\usepackage[utf8x]{inputenc}
\begin{document}

\title{Generation and application of sub-kilohertz oscillatory flows in microchannels}

\author{Giridar Vishwanathan}
\author{Gabriel Juarez}
\thanks{Email address.}\email{gjuarez@illinois.edu}

\affiliation{Department of Mechanical Science and Engineering, University of 
Illinois at Urbana-Champaign, Urbana, Illinois 61801, USA}

\date{\today}


\begin{abstract}

We present a user-friendly and versatile experimental technique that generates 
sub-kilohertz sinusoidal oscillatory flows within microchannels. The method involves 
the direct interfacing of microfluidic tubing with a loudspeaker diaphragm to generate 
oscillatory flow in microchannels with frequencies ranging from $10-1000$ Hz and amplitudes 
ranging from $10 - 600 \ \mu$m. The speaker-based apparatus allows independent control 
of frequency and amplitude that is unique to the speaker's manufacturing specifications. 
The performance of our technique is evaluated by Fourier spectral analysis of oscillatory 
motion of tracer particles, obtained by particle tracking velocimetry, as well as by 
comparing oscillatory flow profiles against theoretical benchmarks such as Stokes flow 
in a square channel and Stokes' second problem near a solid boundary. 
Applications that utilize both the oscillatory flow and the associated steady rectified 
flows are demonstrated in prototypical microfluidic configurations. These include 
inertial focusing and mixing at low Reynolds numbers, respectively.

\end{abstract}

\maketitle

\section{Introduction}

Oscillatory flows in microfluidic devices have been shown to be useful in a range of
applications such as mixing at low Reynolds numbers \cite{Phelan,Ahmed,Frommelt}, 
particle sorting and focusing \cite{Thameem2016,Schmid,Marmottant,Mutlu}, enhancement 
of heat transfer \cite{Qu}, flow control \cite{leslie,phillips}, microrheology 
\cite{VishwanathanPoF,VishwanathanJNNFM}, and chemical extraction \cite{Lestari,Xie}. 
Nevertherless, the widespread use and study of oscillatory flows in microchannels remains 
uncommon due to challenges of implementation. 

At low frequencies ($0.1 \leq f \leq 10$ Hz), oscillatory flows are usually achieved by 
a programmed syringe pump, electromechanical relay valves \cite{Abdolhasani} or a pneumatic 
pressure controller \cite{Zhou}. The fidelity of the desired waveform is limited by inertia 
of the oscillatory driver. For low frequencies, the response time of syringe pumps and 
actuators in electromechanical valves and pnuematic pressure controllers is on the order 
of $\mathcal{O}(10 \ \textrm{ms})$, therefore preventing the realization of sinusoidal 
oscillations at higher frequencies. 

At high frequencies ($10^3 \leq f \leq 10^6$ Hz), piezoelectric transducers, which typically 
possess resonant frequencies in this range, are used \cite{Rallabandi,phillips,Xie,Lieu,Morris}.
The utility of piezoelectric transducers in the 10 - 1000 Hz range are limited by the small 
amplitudes generated. The amplitudes may be partially increased through the use of designed 
features such as membrane cavities in the channel on to which the piezo elements need to be 
bonded to be used properly \cite{Vazquez}. More recent designs of microfluidic oscillators 
primarily aim to achieve oscillatory flows free of external actuators with a focus on 
miniaturization and integration with other lab-on-chip modules. This is typically done by 
exploiting non-linear fluid-elastic interaction with a membrane or diaphragm unit as a steady 
flow is driven through it. Therefore, a time dependent response is obtained even with a steady 
input at low Reynolds numbers \cite{Xia,leslie,Kim,Mosadegh}. Other possibilities such 
as the use of non-Newtonian fluids for switching \cite{Groisman}, generation of oil droplets 
as an oscillatory source \cite{Basilio}, and the Coanda effect \cite{Yang} have also been 
explored. Although these micro-oscillators are highly miniaturized, modular and in 
some cases, capable of producing frequencies in the audible range \cite{Xia}, 
they mostly require the fabrication of complex MEMS devices potentially discouraging 
their use in research attempting to use oscillatory flows. Further, in microfluidic 
oscillators that function based on fluid elastic interaction the amplitudes and frequencies 
are coupled and hence cannot be independently controlled. 

Here, we describe the operation and performance of a simple plug and play apparatus 
capable of producing oscillatory flow in microchannels. This setup allows the 
user to independently control the oscillation frequency in the range of $10-1000$ Hz 
and amplitude in the range of $10-600 \ \mu$m. The aim of this method is simplicity 
and accessibility, hence, allowing researchers to implement oscillatory flow at the 
microscale in a convenient and cost-effective manner without the need for prior 
design constraints and sophisticated microfabrication.

\section{Experimental setup}

\begin{figure*}
\centering
    \includegraphics[width=\linewidth]{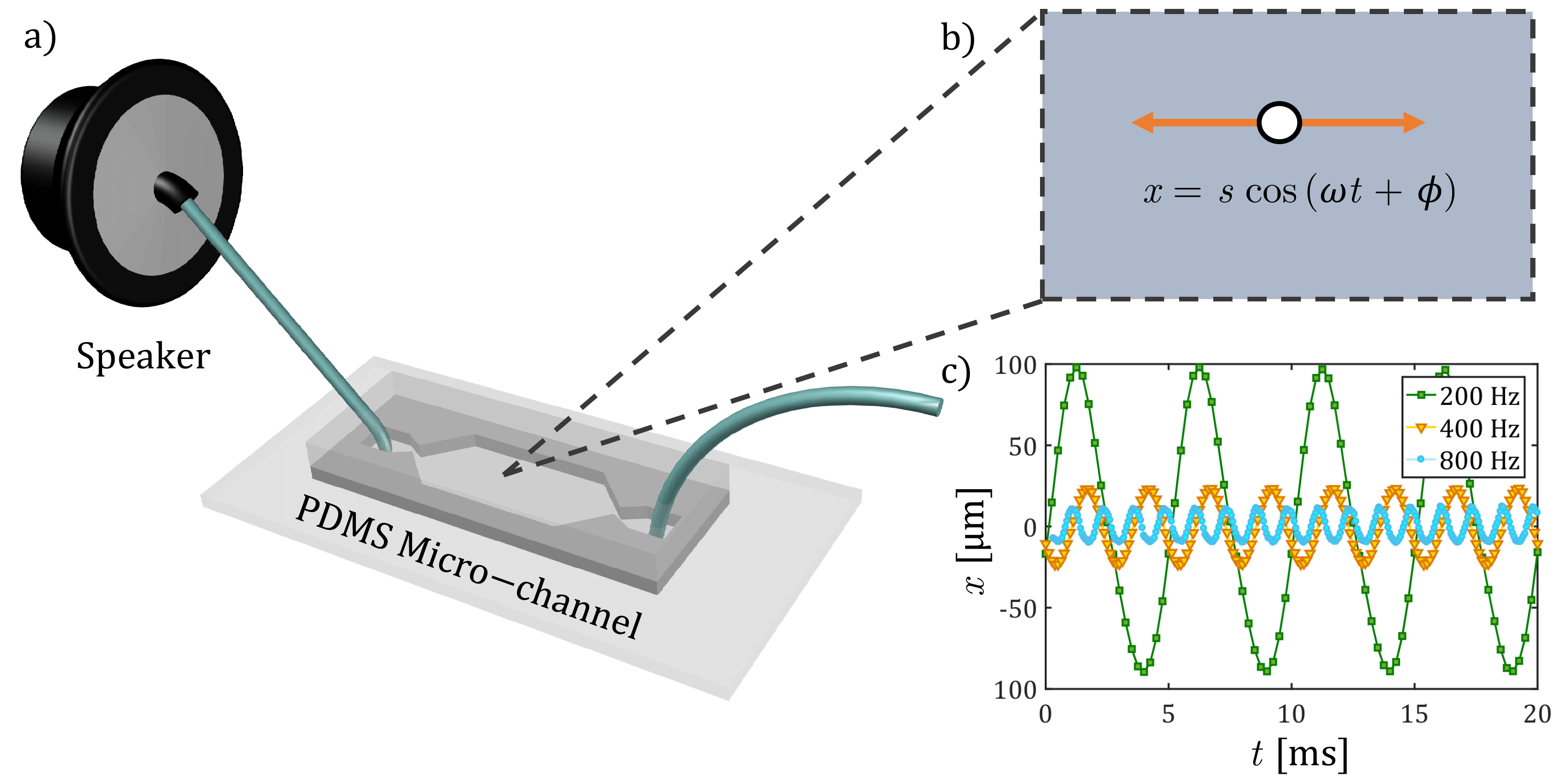}
\caption{
(a) Schematic of the experimental setup. The loudspeaker diaphragm is directly 
interfaced with microfluidic tubing (maintained taut and filled with liquid) to 
generate sinusoidal liquid oscillations in a PDMS microchannel. 
(b) Schematic of tracer particle (and liquid) displacement in a microchannel 
described by the oscillation amplitude and angular frequency. 
(c) Experimental streamwise displacement of $0.93 \ \mu$m diameter tracer particles 
in water over a number of oscillation cycles obtained with micro-particle 
tracking velocimetry.
}
	\label{fig:figone}
\end{figure*}

The apparatus is set up as displayed in the schematic, shown in Figure \ref{fig:figone}(a). 
A PDMS microchannel is bonded to a glass slide and observed through an inverted microscope. 
A loudspeaker (DROK TDA7297B, $15$ W, $90$ dB) is mounted next to the microscope stage. 
The oscillation frequency and amplitude (volume) of the loudspeaker diaphragm are 
controlled by a computer via an auxiliary cable. 
One end of microfluidic tubing (PE60 Intramedic 427416, 0.76 mm ID $\times $ 1.22 mm OD) 
is directly attached to the diaphragm of loudspeaker while the other end is inserted into 
the microchannel outlet. The tubing is maintained taut and its boundary conditions correspond 
to a fixed end at the microchannel outlet and a forced oscillatory displacement at the 
diaphragm. The microchannel and the tubing are filled with liquid during operation. The 
oscillatory displacement of the diaphragm is transduced into elastic deformations of the 
microfluidic tubing at the fixed end of the device outlet. The stress induced by tubing 
deformation generates a time-varying pressure within, resulting in oscillatory displacement
of the liquid at the same frequency ($f$) as the diaphragm. The streamwise displacement 
of a tracer particle in the channel is described by $x=s \cos(\omega t + \phi)$ and 
illustrated in Figure \ref{fig:figone}(b). Here, $s$ oscillation amplitude, $\omega$ is 
the angular frequency, and $\phi$ is the initial phase. 

To characterize the oscillatory flow in microchannels, tracer particles at the midplane 
of the microchannel were observed using brightfield illumination with objectives of 
$10\times$ and $20\times$ magnification (depth of field 8.5 $\mu$m and 5.5 $\mu$m respectively). 
To ensure that the tracers accurately represented the flow, polystyrene tracer particles 
with a mean diameter of $0.93\ \mu$m and density of $1.08$ g/cm$^{3}$ were suspended in 
deionized water (unless mentioned otherwise). The response time associated with the tracer 
particles ($\tau = \rho d^2 /18 \mu \approx 50 \times 10^{-9}$ s) is much smaller than the 
oscillatory timescales considered in this study ($\tau \ll 1/f$). The particle positions 
were recorded using a high-speed scientific CMOS camera with frame rates exactly twenty 
times larger than the driving oscillation frequency ($20f$). The displacement and velocity 
fields are then obtained from 2D particle tracking velocimetry algorithms. 

The microchannel geometries used were the following: (i) a square channel with a width 
and height of $110 \ \mu$m and length of $5$ cm, (ii) a rectangular channel with a  width 
of $5$ mm, height of $200 \ \mu$m, and length of $2$ cm, and (iii) a cross slot channel 
with square cross section with a width and height of $110 \ \mu$m.

The dimensionless groups considered here are the Womersley number, the Reynolds number, 
and the non-dimensional oscillation amplitude. The Womersley number is the ratio of the 
characteristic channel width to the Stokes boundary layer thickness and is defined as 
$\alpha = W \sqrt{ \rho \omega / \mu}$ \cite{Landau}. Here, $W$ is the channel width, 
$\rho$ and $\mu$ are the liquid density and dynamic viscosity, respectively. The Reynolds 
number is the ratio of inertial to viscous forces of the fluid within the microchannel 
and defined as $\textrm{Re}= \rho s\omega W/\mu$ \cite{Thomas2011}. Above, 
$s \omega$ is taken to be the characteristic liquid velocity. The non-dimensional 
oscillation amplitude may be expressed in terms of the Womersley and Reynolds number
as $\epsilon=s/W=\text{Re}/\alpha^2$. For operational conditions considered in this study, 
a microchannel with characteristic width of $200\ \mu$m, and DI water as the working 
liquid, the values of $\alpha$, $\textrm{Re}$, and $\epsilon$ range from 
$1.5 < \alpha < 15$, $0.4 < \textrm{Re} < 80$, and $0.1 < \epsilon < 5$, respectively. 

\section{Results}

\subsection{Oscillatory displacement in microchannels}

Examples of streamwise displacement of individually tracked particles from their mean 
position over a number of cycles during a $20$ ms period are shown in Figure
\ref{fig:figone}(c). The ratio of sampling frequency (camera framerate) to liquid 
oscillation frequency is kept constant at $20$. That is, for oscillation at $200$, $400$ 
and $800$ Hz, a framerate of $4000$, $8000$ and $16000$ Hz is used, respectively. 
The corresponding amplitudes are $100$, $27$ and $14\ \mu$m.

\begin{figure}
\centering
	\includegraphics[width=\linewidth]{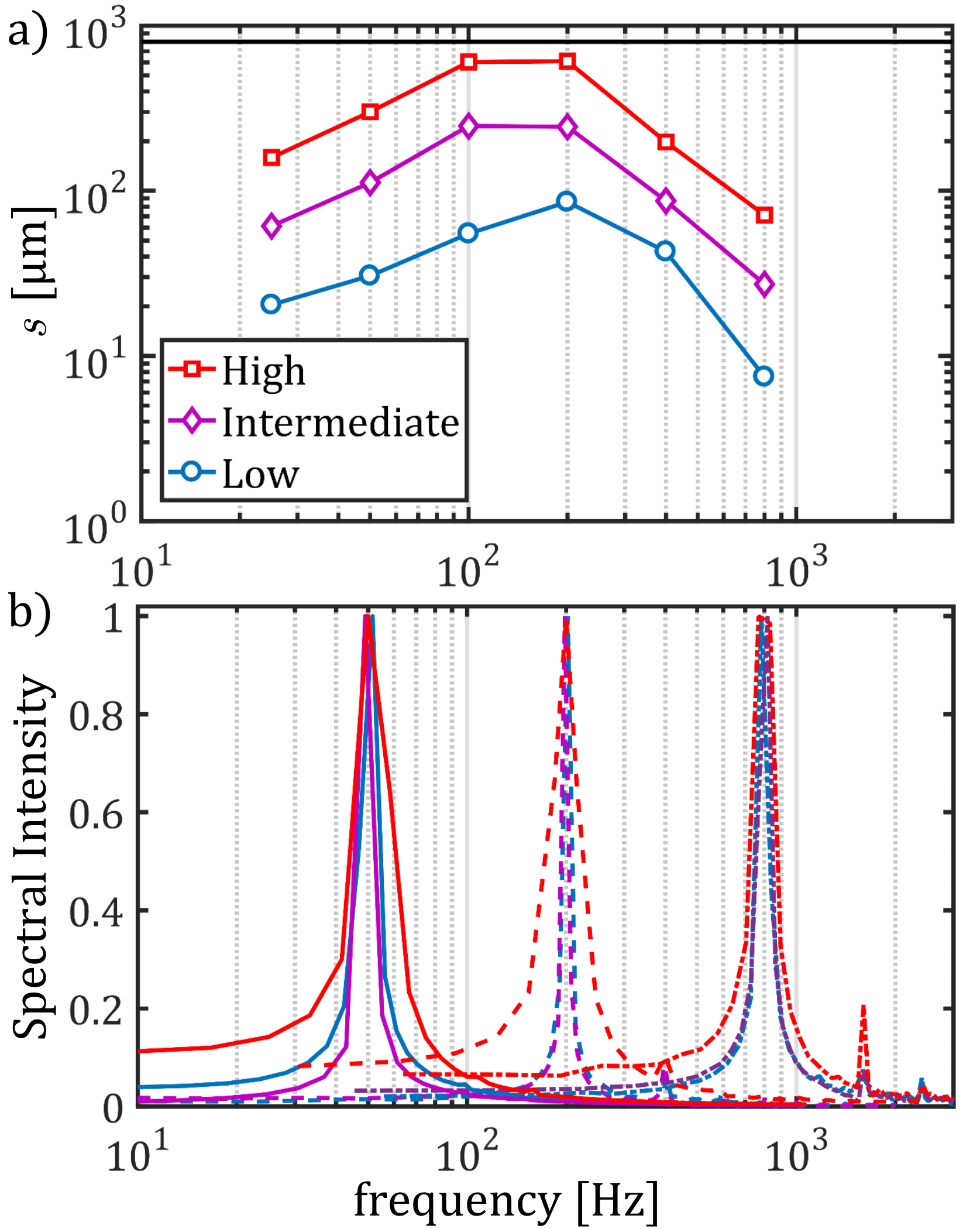}
\caption{
(a) The amplitude of oscillatory displacement in microchannels for a range of 
frequencies and three amplitudes, or speaker volume settings, of low (30\%, blue), 
intermediate (60\%, magenta) and high (90\%, red).
(b) Fourier spectrum analysis of tracer particle displacement in the streamwise 
direction at three different frequencies (50, 200, and 800 Hz) and amplitudes (low, 
intermediate, and high).
}
	\label{fig:figtwo}
\end{figure}

The independent operational range between amplitude and frequency is shown in 
Figure \ref{fig:figtwo}(a). For a given frequency, the displacement of a tracer particle 
is dependent on the volume setting of the loudspeaker. As an example, three volume settings 
are considered here: low (30\%), intermediate (60\%), and high (90\%), where the percentages 
correspond to the maximum speaker volume as determined by the computer. At 100 Hz, for example, 
the amplitude ranges from $50 \ \mu$m at low to $800 \ \mu$m  at high volume setting. The 
amplitude swept by a tracer particle over a single oscillation period, for a given volume 
setting, shows a non-monotonic variation with frequency. Owing to the performance 
characteristics of the speaker, the maximum oscillation amplitude occurs at $200$ Hz, which 
corresponds to the resonant frequency ($\approx 230$ Hz) of the loudspeaker diaphragm. The 
horizontal black line indicates the maximum particle oscillation amplitude of $800 \ \mu$m 
that can be measured due to the field of view limited by the camera when using a $10\times$ 
microscope objective lens. 

A Fourier spectrum analysis of particle trajectories at varying oscillation amplitudes and 
frequencies is shown in Figure \ref{fig:figtwo}(b). The spectra have been obtained for 
oscillation frequencies of $50$, $200$ and $800$ Hz and at volume settings of low, 
intermediate, and high. For low and intermediate volume settings, monodisperse peaks in the 
spectral intensity correspond to the input driving frequency of the loudspeaker. The peaks 
are especially narrow at $200$ Hz, or near the resonance frequency of the diaphragm. For high 
volume settings, the peaks correspond to the input driving frequency, however, widening of the 
peak is noticeable. In some cases, such as for $800$ Hz at high volume setting, contributions 
due to higher harmonics are of considerable strength.

A quantitative measure of harmonic distortion present in the signal as compared to the 
fundamental driving frequency is obtained by calculating the total harmonic distortion (THD). 
The THD is defined as $\textrm{THD} =\sqrt{\Sigma^{N}_{i=1} V_i^2}/V_1$, where $V_i$ is the 
power of the spectral intensity at the $i$th harmonic \cite{Shmilovitz}. A low THD value is
associated with a more accurate representation of the original driving signal. For low volume
settings, the THD at $50$, $200$, and $800$ Hz are $3.5\%$, $7.1\%$, and $9.1\%$, respectively. 
For intermediate volume settings, the THD at $50$, $200$, and $800$ Hz are $5.2\%$, $8.9\%$, 
and $13.4\%$, respectively.For high volume settings, the THD at $50$, $200$, and $800$ Hz are
$7.3\%$, $11.1\%$, and $21.3\%$, respectively. The growing magnitude of higher harmonics with
increasing speaker volume typically limits operation at frequencies $>400$ Hz to low or
intermediate speaker volumes. At low frequencies however, the maximum amplitude is chosen 
to avoid damage to the microchannel or unfastening of the outlet tube from the speaker cone.
Therefore, sinusoidal oscillations with amplitudes ranging from $10 < s < 200 \ \mu$m can 
be reliably achieved throughout the entire range of frequency.

The maximum pressure inside the square channel may be estimated from the modified Poiseuille 
formula:
\begin{equation} \label{eq:pressuredrop}
    \Delta P=\frac{64\mu Lsf}{D_h^2} \ ,
\end{equation}
where $D_h$ and $L$ are the hydraulic diameter and length of the channel. For the square 
channel ($D_h=110\ \mu$m and $L=5$ cm) filled with DI water ($\mu=1.002$ mPa s) and settings 
for maximum oscillatory displacement ($f=200$ Hz, $s=600 \ \mu$m), the pressure inside the 
channel is calculated to be approximately equal to $31$ kPa.

\subsection{Oscillatory flow in microchannels}

\begin{figure}
\centering
	\includegraphics[width=\linewidth]{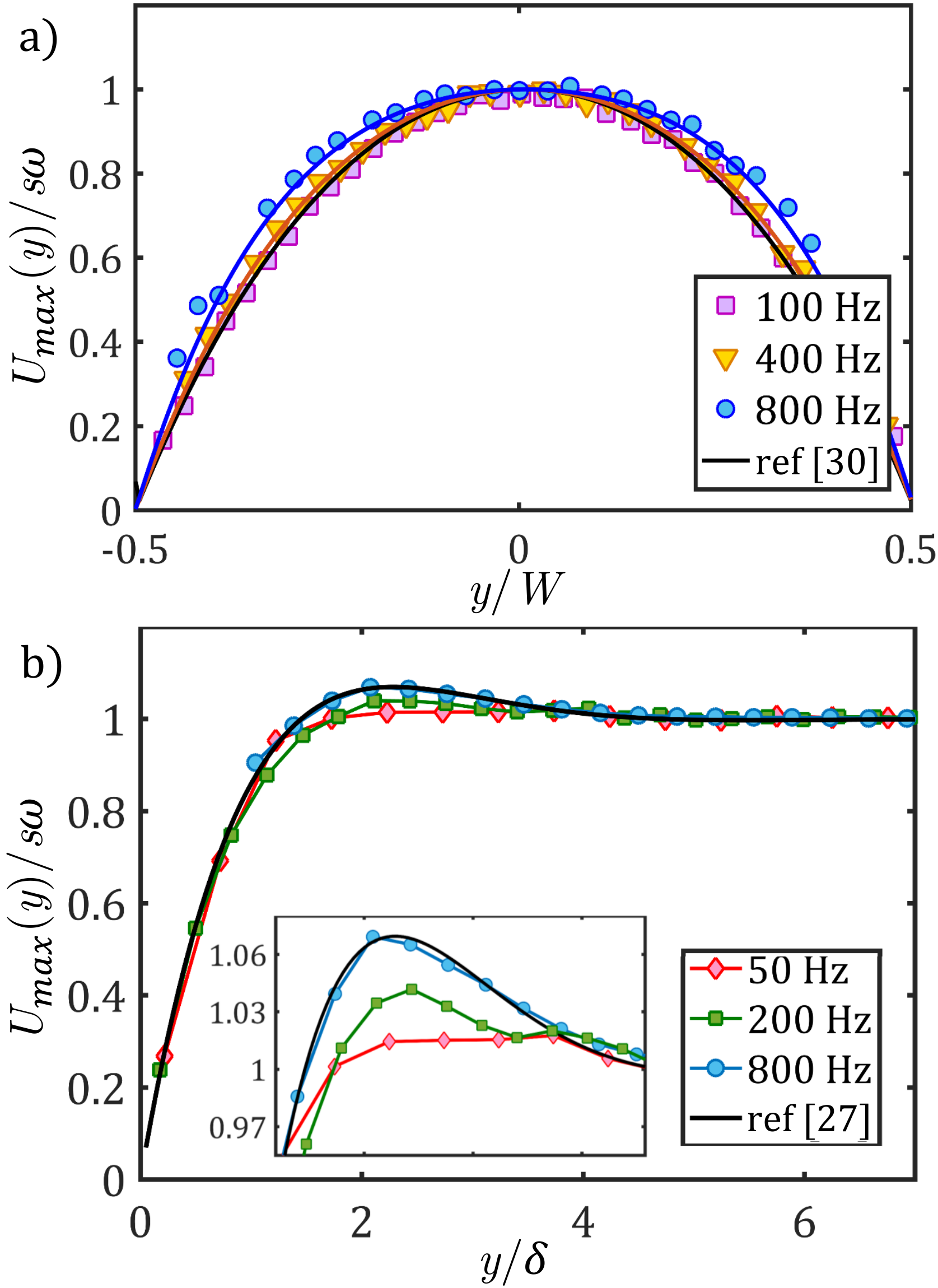}
\caption{
(a) Oscillatory velocity flow profile in a square channel (symbols) showing deviation 
from Stokes flow profile (black solid curve) with increasing frequency.
(b) Oscillatory velocity profile with position (symbols) near a solid channel wall in a 
semi-infinite rectangular channel ($W \gg H$) showing agreement with the theoretical 
solution (black solid curve) to Stokes' second problem \cite{Landau} with increasing 
frequency. (Inset) Close-up of the region from $1.5 \leq y/\delta \leq  4.5$ to demonstrate
agreement with equation (\ref{eq:stokes2ndprob}) at different frequencies.
}
	\label{fig:figthree}
\end{figure}

The small length scales of $\mathcal{O}(100 \ \mu \textrm{m})$ associated with microchannels 
imply that most microscale flows are laminar flows governed by the Stokes equation. An 
important feature of microscale oscillatory flows in the $10-1000$ Hz range is that transient 
effects associated with the unsteady Stokes equation become significant. An example of departure 
from Stokes flow is illustrated by the comparison of the steady Stokes flow velocity profile
\cite{Obrien} at the midplane (black solid curve) against those obtained experimentally for 
oscillatory flow at different frequencies (symbols), shown in Figure \ref{fig:figthree}(a).

To obtain the amplitude of velocity in the square channel midplane, $50-200$ particles 
($0.93 \ \mu$m diameter) are tracked for one hundred oscillation cycles and their respective
velocities are computed. The amplitude of each velocity series ($U_{max}(y)$) is obtained and
superposed in the streamwise direction. The resulting spread of speeds is filtered for outliers 
and averaged. The associated statistical error bars are smaller than the data markers shown. 

The results for $100$ Hz and $400$ Hz are similar to the Stokes laminar flow profile (black 
solid curve). At $800$ Hz, however, there is considerable deviation from the steady velocity 
profile due to increasing $\alpha$ with frequency. For the cases of $100$, $400$, and $800$ Hz 
the $\alpha$ values are $2.75$, $5.51$, and $7.78$, respectively. For these $\alpha$ values, 
the analytical series solution for the amplitude of the midplane oscillatory velocity profile 
($U_{max}(y)$) was evaluated correct to one hundred terms \cite{Obrien}, and are shown by the
continuous lines to good agreement with experimental data, even at $800$ Hz.


In contrast, deviations from the unsteady Stokes equation are demonstrated in Figure
\ref{fig:figthree}(b) by comparing the amplitude of measured oscillatory flow (symbols) in 
the rectangular channel ($W \gg H$), with those obtained theoretically from the solution to 
oscillatory flow over an infinite flat plate (Stokes' second problem) \cite{Landau, Wang89}. 
The maximum temporal flow velocity as a function of distance from a flat plate is given by 
the expression:
\begin{equation} \label{eq:stokes2ndprob}
    U_{max}(y)/s\omega=\sqrt{2} e^{-y/(2\delta)} \sqrt{\cosh{(y/\delta)} -\cos{(y/\delta)}} \ ,
\end{equation}
where $\delta$ is the Stokes boundary layer length and equal to $\sqrt{\mu / \rho \omega}$.

At $800$ Hz, normalized experimental data (symbols) is in good agreement with equation 
(\ref{eq:stokes2ndprob}). At $50$ Hz, and to a lesser extent at $200$ Hz,  deviations from the 
theory occur where the velocity amplitudes are larger than those at the far field and is 
detailed in the inset. This is due to the relatively short channel height ($200\ \mu$m) and 
the effect of boundary layers at the top and bottom walls of the channel affecting flow at 
the midplane. The corresponding values of $\alpha$ for the cases of $50$, $200$, and $800$ 
Hz are $3.57$, $7.14$, and $14.3$, respectively. Therefore, at the midplane, effects of 
channel side-walls may be neglected for distances larger than $4\delta$ into the channel. 
Further, three dimensional flow effects can be ignored for $\alpha \geq 7.5$ when the 
shorter dimension is used.

\subsection{Applications of oscillatory flow in microchannels}

\begin{figure*}
\centering
	\includegraphics[width=\linewidth]{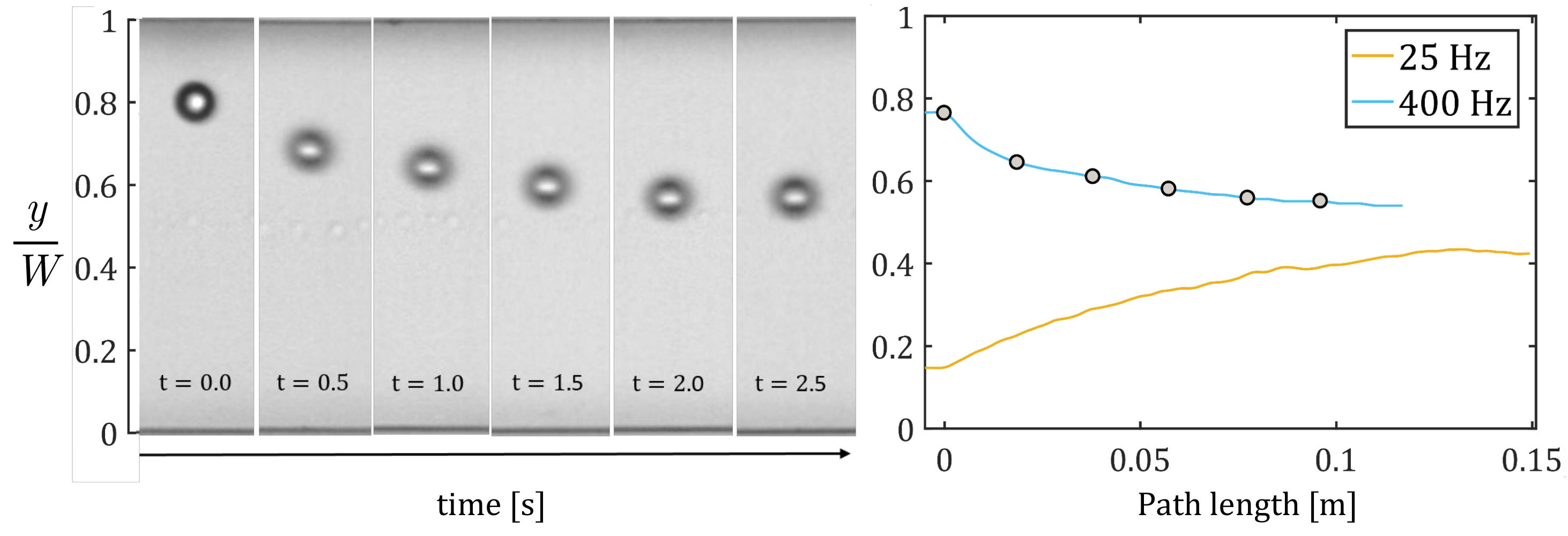}
\caption{
(a) Inertial focusing of a $10 \ \mu$m polystyrene particle in oscillatory flow with 
a frequency of $400$ Hz in a $110 \ \mu$m square channel. 
(b) The spanwise location of the particle centroids as they migrate to the equilibrium 
position at the center of the channel during oscillatory flow.
}
	\label{fig:figfour}
\end{figure*}

The application of oscillatory flows in microchannels may be broadly divided into two 
categories. The first category of flows are those where oscillatory motion enables 
instantaneous local velocities or shear rates without net displacement, which is usually 
implemented to reduce device footprint and allow for prolonged observation 
\cite{Jo,Alizadehgiashi,Mutlu,Abdolhasani}. The second category of flows are those that 
utilize steady rectified flows associated with an underlying primary high frequency 
oscillatory flow \cite{Riley}, which have been shown to be useful in mixing \cite{Ahmed}, 
hydrodynamic manipulation of particles and cells \cite{Lutz,Lieu,Thameem2016}, and more 
recently, in microrheology \cite{VishwanathanPoF,VishwanathanJNNFM}. Here, we 
demonstrate two specific applications of the oscillatory driver from each category, 
namely, inertial focusing from the former and mixing from the latter, using simple 
prototypical microfluidic configurations. 

\subsubsection{Inertial focusing}

Inertial focusing in microchannels is a passive technique where suspended particles 
undergoing unidirectional flow migrate across streamlines, due to particle inertia, to 
equilibrium focus positions \cite{DiCarlo,DiCarlo2009,Martel,Stoecklein}. In straight 
channels, of rectangular or circular cross section, the competing forces that lead to 
particle migration are the wall interaction force, which directs the particle away from 
the channel wall, and the shear gradient force, which directs the particle toward the 
channel wall. The summation of these forces is termed the inertial lift and the 
equilibrium position of the particle is determined once the opposing forces are balanced. 
Factors influencing the inertial lift force are the channel geometry, flow rate, and 
particle size. 

Since it is a high-throughput method for non-contact manipulation at the microscale, 
inertial focusing has been utilized in numerous applications ranging from flow 
cytometry \cite{Hur,Bhagat,Oakey}, size sorting \cite{Kuntaegowdanahalli,Wu,Nivedita},
mixing \cite{Amini}, and filtration \cite{Seo}. An important parameter when designing 
channels for applications is the length required to reach the equilibrium focus position, 
estimated \cite{DiCarlo2009} to be equal to
\begin{equation} \label{eq:focuslength}
    L_f = \frac{\pi \mu D_h^2}{ \rho U_m a^2 C_{\ell}}\ ,
\end{equation}
where $U_m$ is the maximum flow velocity and $C_{\ell}$ is the lift coefficient, which 
typically varies in the range of $0.02 - 0.05$. From this relation, 
it is apparent that sufficiently high velocities and large particles are required to 
minimize the focusing length. Recent work, however, demonstrates the use of oscillatory 
flows for inertial focusing, where oscillatory flow at relatively low frequencies 
($<20$ Hz) results in a channel of practically infinite length \cite{Mutlu}. Thus, 
it becomes possible to focus particles with far smaller particle Reynolds numbers 
corresponding to $\textrm{Re}_{\textrm{p}}= (a/W)^2 \textrm{Re} <0.01$. 

\begin{figure*}
\centering
	\includegraphics[width=\linewidth]{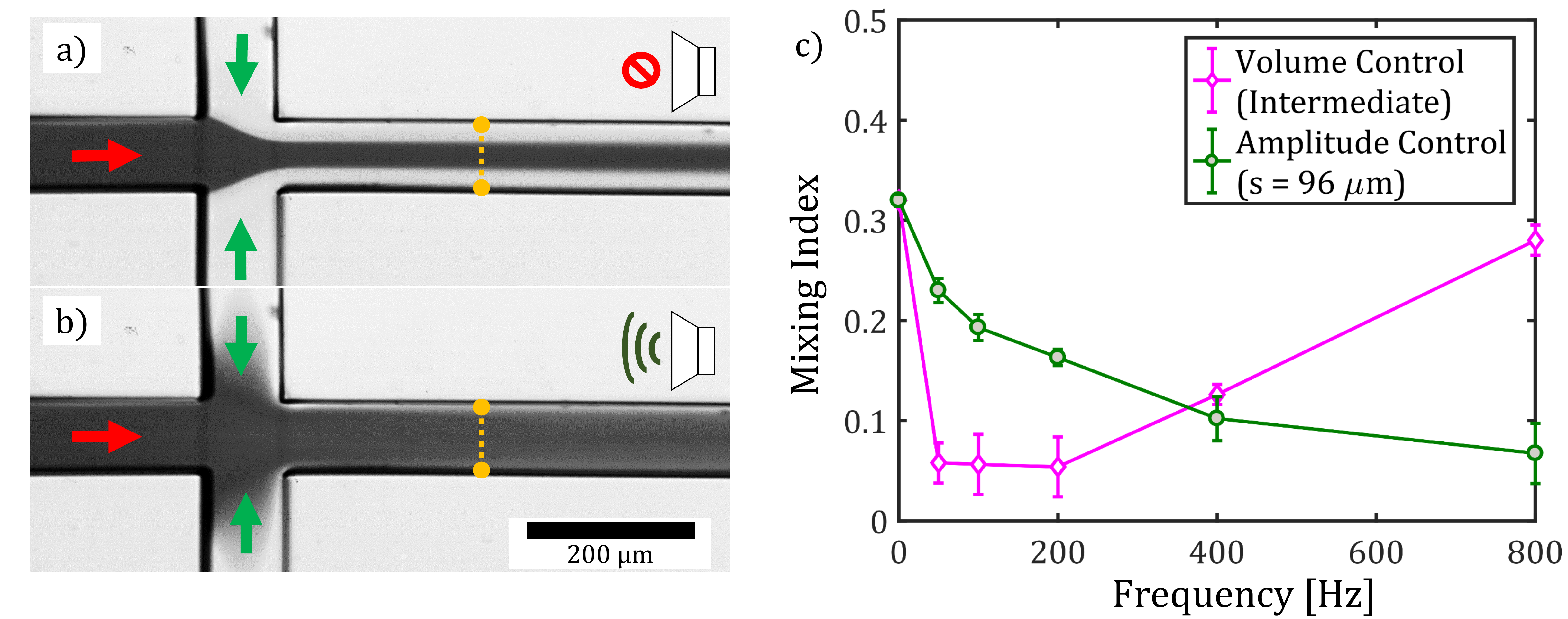}
\caption{
(a) Mixing of co-flowing streams of $30\%$ (w/w) aqueous glycerol solutions without
oscillations applied. 
(b) Mixing of co-flowing streams of $30\%$ (w/w) aqueous glycerol solutions with 
oscillations applied at a frequency of $400$ Hz.
(c) Mixing index as a function of frequency for constant amplitude and constant 
volume settings measured at a distance $2.5W$ downstream from the cross-slot 
region.
Scale bar is $200 \ \mu$m.
}
	\label{fig:figfive}
\end{figure*}

Inertial focusing of a $10\ \mu$m polystyrene particle in oscillatory flow of $400$ Hz 
with an amplitude of $22\ \mu$m is demonstrated in the square channel, for which,
$\textrm{Re}_{\textrm{p}}=0.050$ and $a/W=0.091$. Micrographs of a single polystyrene 
particle at regular time intervals as it migrates to the equilibrium position is shown 
in Figure \ref{fig:figfour}(a). Using stroboscopic imaging, the lateral migration toward 
the center of the channel is apparent. The corresponding vertical position of the particle 
centroid, as determined by particle tracking, is shown as a function of the approximate 
path length traversed by the particle in Figure \ref{fig:figfour}(b). The marked points 
(symbols) correspond to the instances shown in the micrographs. The path length is 
estimated as $4sft$, where $4s$ is the distance covered by the particle at the center 
of the channel in a single oscillation cycle and $t$ is the time elapsed after the start 
of oscillations. The focusing behavior of another particle at $25$ Hz and an amplitude 
of $36\ \mu$m is also shown. The corresponding path length for focusing (or focus length) 
is found to be about $0.15$ m, comparable to the $400$ Hz case and in good agreement 
with previous results for similar conditions ($\textrm{Re}_{\textrm{p}}=0.050$ and 
$a/W=0.125$) \cite{Mutlu}. The time required for focusing, however, is about $3$ 
seconds for the $400$ Hz case and $35$ seconds for the $25$ Hz case.  

The advantages of oscillatory flow for inertial focusing include decreased channel 
lengths, lower pressure drops, and lower shear rates. Because there is no net 
displacement, the particle remains in the field of view as it migrates to the 
equilibrium position. In contrast with unidirectional flow, the approximate channel 
length required for the particle to reach it's equilibrium position as determined 
by equation (\ref{eq:focuslength}), where $U_m$ is given by the characteristic 
fluid velocity $s \omega$, is $L_f \approx 1.37$ m, which is impractical. 
The lower pressure drop allows for the convenient fabrication and use of PDMS 
microchannels. The combination of lower pressure drop and lower shear rates is 
of particular interest in biomedical application where cells are susceptible to 
damage induced by fluid stresses. 

\subsubsection{Microscale mixing}

Low Reynolds number ($\textrm{Re}\leq1$) flow in microchannels present a significant 
challenge to applications involving mixing \cite{Ottino}. This is because, the dominant 
mechanism of mixing is diffusion in the absence of chaotic advection, which is normally 
associated with high $\textrm{Re}$ mixing. The effectiveness of mixing is quantified by 
estimating the length required to achieve mixing ($L_m$). For purely diffusive mixing,
\begin{equation} \label{eq:mixlength}
    L_m \geq D_h \textrm{Pe} \ .
\end{equation}
Above, $\textrm{Pe}$ is the Peclet number and defined as the ratio of $U_aW/D$, where 
$U_a$ is the average flow velocity and $D$ is the diffusion coefficient. For 
$\textrm{Re} \leq 1$ and diffusion coefficients of $D \approx 100 \ \mu$m$^2$ s$^{1}$, 
the Peclet number would be $\textrm{Pe} \geq 10^4$, resulting in a channel length for 
sufficient mixing to be $L_m \geq 1$ m, which is undesirable.

To overcome this challenge, a variety of microscale mixers have been developed to enhance 
mixing at low Re, and are categorized as either passive or active mixers
\cite{Hessel,Nguyen,Lee,Ward,lee16,Cai}. Passive mixers make use of the channel geometry, 
usually incorporating repeating complex or 3D channel features to enhance mixing of two 
streams flowing together at a constant rate. On the other hand, active mixers rely on 
externally applied forces and are further categorized based on the nature of the external
actuation \cite{Ober}. One such category of mixers are acoustic micromixers 
\cite{Liu,Ahmed,Bachman} that rely on external acoustic actuation to generate steady 
rectified flows that mix different liquids.

Here, mixing of two aqueous glycerol solutions ($30\%$ w/w, $\mu=2$ mPa s), one with 
colored dye and one without, is demonstrated in the cross-slot channel using steady 
rectified flows. As seen from figure \ref{fig:figfive}(a), the dyed solution enters the 
cross-slot with a flow rate of $0.3\ \mu$L/min (red arrow) while the undyed solution 
enters the cross-slot from either side with identical flow rates of $0.15 \ \mu$L/min 
(green arrows). This results in a total flow rate of $0.6 \ \mu$L/min at the outlet. 

When no oscillatory flow is imposed, minimal diffusive mixing at the interface is observed 
within the field of view. When oscillatory flow of $400$ Hz is imposed on the same 
configuration, steady vortices are generated near the corners of the cross-slot as seen 
in Figure \ref{fig:figfive}(b), which facilitate mass transfer across the interface 
through advection. A large exposure time of $20$ ms was used for imaging so that 
variations over signal phase are averaged.

Mixing performance is quantified by first obtaining the intensity profile across the 
channel at a distance $2.5W$ downstream from the center of the cross slot, indicated 
by the yellow dashed lines in Figure \ref{fig:figfive}(a) and (b). The standard 
deviation of the mixture fraction profile, derived from the intensity values, is used 
as the mixing index and is defined as $\textrm{MI} = \sqrt{\Sigma(I_i - I_m)^2/N}$, 
where, $I_i$ is the pixel intensity value and $I_m$ is the pixel intensity of a completely 
mixed solution, and $N$ is the number of sampling points \cite{Liu00}. The values of 
the index range from $\textrm{MI}=0.5$ for completely unmixed to $\textrm{MI}=0$ 
for completely mixed solutions. A value of $\textrm{MI} \leq 0.1$ indicates sufficient 
mixing. 

The variation of the mixing index as a function of frequency for both constant 
oscillation amplitude and constant volume setting are shown in Figure \ref{fig:figfive}(c).
For constant amplitude settings, the mixing index decreases monotonically with increasing 
frequency, implying sufficient mixing for $f \geq 400$ Hz. The improved mixing with 
increasing frequency is due to the increase in magnitude of the steady rectified flow 
velocities, which scale as $\mathcal{O}(s^2\omega)$. For a constant volume (intermediate) 
setting, however, the mixing index is non-monotonic, with sufficient mixing occurring in 
the range of $100-200$ Hz. Based on the amplitude characteristics, shown in Figure
\ref{fig:figtwo}(a), for constant volume settings, the largest amplitudes occur near 
the resonance frequency of the loudspeaker diaphragm. 

For the specific case presented here, the length demonstrated to achieve good mixing with 
steady rectified flows ($L_m \leq 250 \ \mu$m) is much less than the length required for 
mixing according to equation (\ref{eq:mixlength}), which is calculated to be $L_m = 2.5$ cm. 
Although the nature of forces involved are purely hydrodynamic, oscillatory flows with 
independently controllable amplitude and frequency allow for the decoupling of flow rate 
from the rate of mixing which is not possible for passive micromixers. Additionally for 
the range of frequencies and amplitudes achieved here, strong rectified flows are achieved 
near solid boundaries in the microchannel as opposed to the boundaries of bubbles used with 
ultrasonic frequencies elsewhere \cite{Liu,Ahmed} which are unstable at long operation times. 
Lastly, the implementation of this method in combination with any other passive or active 
technique can further enhance mixing at the microscale by increasing the number of passes 
without affecting flow rate.

\section{Discussion}

We have discussed an accessible, effective, and versatile plug and play technique to generate
oscillatory flows over a range of amplitudes and frequencies in microchannels. By directly
interfacing microfluidic tubing with a loudspeaker diaphragm, sub-kilohertz oscillatory 
flows in the range of $10-1000$ Hz with amplitudes in the range of $10-600 \ \mu$m are 
produced. The corresponding wavelengths lie between $1-100$ m and are far larger than 
the dimensions of a typical microchannel. Thus, nearly unattenuated flows of a uniform 
phase can be achieved throughout. This is in contrast with flows in the $10^4-10^7$ Hz 
range where attenuation is significant and effects are usually local to the transducer.
The resulting velocity profiles can also be tuned from Stokes-like flow ($\alpha\leq4$) 
at low frequencies to plug-like unsteady Stokes flow ($\alpha\geq7.5$) at high frequencies 
allowing for manipulation of the flow profile for a given micro-scale geometry. In addition 
to coherent oscillatory flows, strong and well defined rectified flows near curved boundaries 
and interfaces are also made possible in this frequency and amplitude range.

The guiding principles for applications are two-fold. First, oscillatory flows permit an 
increase the net distance travelled by the fluid without an accompanied increase in flow 
rate, shear rate, pressure drop, or particle displacement seen in long channel steady flows 
and high throughput applications. More generally, such conditions are particularly useful 
when the desired sample or analyte response to the flow environment is directionally invariant. 
In theory, the resulting effect can be increased indefinitely by simply increasing the frequency 
of oscillation. Although in practice, attenuation and secondary flows which grow stronger with
increasing frequency prevent this from being realized. Owing to the large wavelengths in this 
range of flows, attenuation is limited while secondary flows usually become significant only 
for frequencies larger than 100 Hz implying that an optimum frequency is likely encountered 
in the range of frequencies realized here. The resulting lower steady pressure drop and steady 
shear rates are particularly useful for cells and other suspended biological matter that 
are sensitive to or potentially damaged by prolonged exposure to excessive shear rates or 
pressures. The absence of particle net displacement is useful in decreasing the footprint 
of microfluidic devices and in situations where continuous observation is needed to track 
the evolution in processes such as cell growth or chemical synthesis in dynamic environments.

Second, the steady rectified flow speeds are of considerable magnitude despite the maximum 
frequency considered being much smaller than the typical resonant frequencies of 
piezoelectric transducers ($1-100$ kHz). This is because of the large amplitudes accessed 
by the loudspeaker diaphragm and the dependency of flow speed on amplitude, which scales 
as $s^2\omega$. Further, the Stokes boundary layer thickness is accurately controlled and 
varies in size from $10 \leq \delta \leq 100 \ \mu$m for the highest and lowest frequencies 
accessed here, respectively. Therefore, the flow pathlines are less sensitive to 
manufacturing defects and feature surface quality, making this approach more amenable to 
precision applications such as sorting, trapping and manipulation of particles and cells,
viscometry, and other controlled mass transfer applications. 

\bibliography{oscflows}

\end{document}